\documentclass[aps,showpacs,prb,twocolumn]{revtex4}

\usepackage{graphicx}
\usepackage{dcolumn}
\usepackage{bm}

\begin{document}


\title{Evidence for $\pm$s-wave pairing symmetry in LiFeAs: specific heat study}

\author{Dong-Jin Jang, J. B. Hong, Y. S. Kwon,  and T. Park}
\email{tp8701@skku.edu}
\affiliation{$^1$Department of Physics, Sungkyunkwan University,
Suwon 440-746, Republic of Korea}
\author{K. Gofryk, F. Ronning, J. D. Thompson}
\affiliation{Los Alamos National Laboratory, Los Alamos, NM 87545, USA}
\author{Yunkyu Bang}
\email{ykbang@chonnam.co.kr}
\affiliation{ Department of Physics, Chonnam National University, Kwangju 500-757, Republic of Korea}


\begin{abstract}
We report specific heat capacity measurements on a LiFeAs single crystal at temperatures down to 400~mK and magnetic fields up to 9~Tesla. A small specific heat jump at $T_c$ and finite residual density of states at $T=0$~K in the superconducting (SC) state indicate that there are strong unitary scatterers that lead to states within the SC gap.
A sub-linear magnetic field dependence of the Sommerfeld coefficient $\gamma(H)$ at $T=0$~K is equally well fitted by both a nodal d-wave gap as well as a sign changing multiband $\pm$s-wave gap. When impurity effects are taken into account, however, the linear temperature dependence of the electronic specific heat $C_{el}/T$ at low temperatures argues against a nodal d-wave superconducting gap. We conclude that the SC state of LiFeAs is most compatible with the multiband $\pm$s-wave SC state with the gap values $\Delta_{small}=0.46\Delta_{large}$.
\end{abstract}
\maketitle

\section{Introduction}
Information on the superconducting (SC) pairing symmetry is important for the understanding of the SC pairing mechanism. For the newly discovered Fe-pnictide superconductors, determination of the SC order parameter has been controversial partly due to its sensitive dependence on measurement probes and material stoichiometry.~\cite{Kamihara,Ishida} In the '1111' phase of REFeAsO (RE = rare earth), a fullly gapped scenario is supported by angle-resolved photoemission spectroscopy(ARPES) and tunnelling measurements,~\cite{Kondo, Chen} but a gap symmetry with nodes on the Fermi surface is suggested by far-infrared ellipsometric measurements.~\cite{Dubroka} Penetration depth measurements reported either a full gap or a nodal gap for RE = Sm,Pr~\cite{Malone,Hashimoto} and RE=La,~\cite{fletcher09} respectively. Identification of the SC gap symmetry is also
controversial in the '122' phase of AeFe$_2$As$_2$ (Ae = alkaline earth). For electron doped BaFe$_2$As$_2$, full gaps are reported from ARPES~\cite{Terashima} and scanning tunnelling microscopy (STM) measurements,~\cite{Yin} but anisotropic or nodal gaps are supported by specific heat,~\cite{Jang} thermal conductivity,~\cite{Tanatar,Dong} penetration depth,~\cite{Gordon,Hashimoto} Raman scattering,~\cite{Mushuler} nuclear magnetic resonance (NMR),~\cite{Ning} and muon spin resonance ($\mu$SR).~\cite{Williams} For hole-doped Ba122, multiple gaps without nodes are reported from specific heat~\cite{Mu} and thermal conductivity.~\cite{Luo}

LiFeAs, the so-called '111' phase, is unique in that it is superconducting without carrier doping and the residual resistivity ratio (RRR) is about 50, one of the highest among Fe-based SC compounds,~\cite{song10} holding promise for a determination of the superconducting gap symmetry without complications from material defects.~\cite{stewart09,Gofryk} Various spectroscopic and thermodynamic measurements suggest an isotropic SC gap without nodes on the Fermi surface.~\cite{Inosov, Wei, kim11, tanatar11} Previous specific heat measurements, a direct bulk probe of the electronic density of states (DOS), suggest multiple SC gaps to explain its temperature dependence.~\cite{Wei,stockert11} The analysis, however, is limited due to lack of specific heat data below 2~K, where a difference in the DOS becomes prominant among different types of SC pairing symmetry. In order to elucidate the nature of the order parameter of LiFeAs, we measured the specific heat of LiFeAs single crystals down to 400~mK and under magnetic field up to 9~Tesla. A large value of the Sommerfeld coefficient $\gamma_0 (\approx 7.5$~mJ/mol$\cdot$K$^2$) in the SC state at zero magnetic field and small specific heat jump ratio $\Delta C / C_{n}$ at $T_c (\approx 0.5)$ indicates the presence of large elastic scattering due to impurities. Here $C_{n}$ is the specific heat at $T_c$ in the normal state. When the impurity effects are taken into account, observation of a linear temperature dependence of the low-$T$ specific heat $C/T$ and a sublinear magnetic field dependence of the Sommerfeld coefficient $\gamma (H)$ excludes a d-wave SC gap, but is consistent with the sign changing multiband $\pm$s-wave state with the gap values $\Delta_{small}=0.46\Delta_{large}$.

\section{Experiments}
Single crystalline LiFeAs, which is formed in a $P4/$nmm tetragonal Cu$_2$Sb-type structure with $a=3.7818~\AA$ and $c=6.3463~\AA$, was synthesized in a sealed tungsten crucible by the Bridgeman method.~\cite{song10} X-ray diffraction pattern analysis
showed that crystals from this batch are homogeneous and well
oriented. This growth technique seems to avoid inclusion of
impurities that can lead to an anomalous Schottky-like upturn in
the low-temperature specific heat of Fe-based superconductors,
making it a useful synthesis technique for a specific heat study
of SC gap symmetry.~\cite{Jang} The crystals used in the current
study are from the same batch that has been used for previous
transport, thermodynamic, and spectroscopic studies.~\cite{song10,
tanatar11, kim11} A Quantum Design PPMS (Physical Properties
Measurement System) with a $^3$He option was used to measure
specific heat down to 400~mK and up to 9~T for magnetic fields applied parallel and perpendicular to the crystalline c-axis.

\begin{figure}[tbp]
\includegraphics[width=7.5cm]{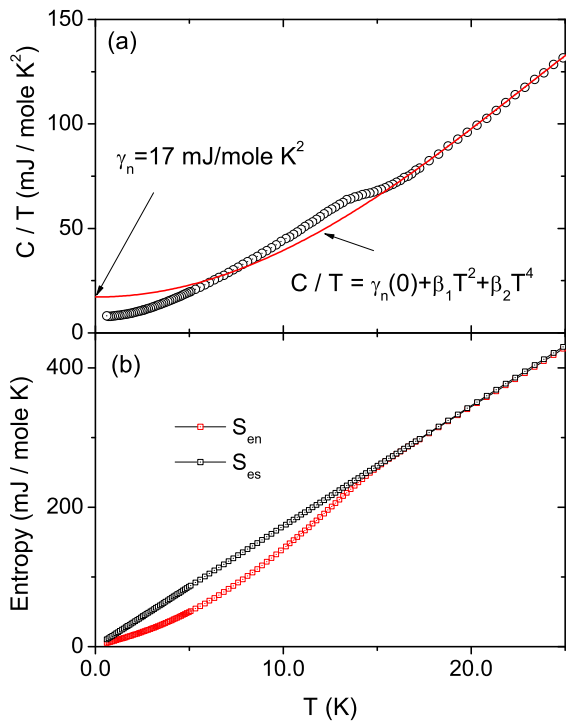}
\caption{(color online) (a) Specific heat of LiFeAs at zero
magnetic field is plotted down to 400~mK. The red solid line is an
estimated normal state specific heat $C_{n}/T$ from a polynomial
form described in the text. (b) Calculated electronic entropy in
the superconducting state ($S_{es}$) and in the normal state
$S_{en}$.} \label{fig1}
\end{figure}

\section{Results}
Figure 1(a) shows the temperature-dependent specific heat capacity
of a LiFeAs single crystal at zero field. A superconducting transition
occurs at 15~K (=$T_c$), which is determined from the mid-point of the
the specific heat anomaly. The measured specific heat ($C$)
consists of electronic ($C_{el}$) and phononic ($C_{ph}$)
contributions, i.e., $C=C_{el}+C_{ph}$. Non-Debye behavior has
been often reported in Fe-based superconductors and ascribed to a
large Einstein contribution or a low Debye temperature $T_D$.
Because an Einstein phonon contribution is negligible below 25~K in
LiFeAs, a second term of the harmonic-lattice approximation is
added to the phonon specific heat:
$C_{n}=\gamma_n(0)T+\beta_1T^3+\beta_2T^5$.~\cite{stockert11} The
specific heat is best described by the least-squares fit with
$\gamma_n(0)=17.0\pm0.9$~mJ/mole$\cdot$K$^2$,
$\beta_1=0.231\pm0.004$~mJ/mole$\cdot$K$^3$ and
$\beta_2\simeq-0.0001$~mJ/mole$\cdot$K$^5$. The electronic
Sommerfeld coefficient
$\gamma_n(0)$($=17.0\pm0.9$~mJ/mole$\cdot$K$^2$) at zero magnetic
field in the normal state is comparable with previously reported
values of 23, 20, and 10~mJ/mole$\cdot$K$^2$ from
Refs.~[28], [24], and~[27], respectively. The Debye temperature estimated from the fit is
294~K, which is similar to 310~K from Ref.~[27]. Figure~1b shows the electronic entropy in the normal (squares) and superconducting (circles) states after
subtracting the non-electronic contribution. The two entropies
become equal at $T_c$, which satisfies the entropy constraint,
$0=\int_0^{T_c}(C-C_{n})/TdT$. From the electronic entropy in
Fig.~1(b), we estimate the superconducting condensation energy
$U=\int_{0}^{T_c}(S_{en}-S_{es})dT=B_c(0)^2/2\mu_0\simeq$363~mJ/mole
and thermodynamic critical field $B_c(0)\simeq$0.23~T. When we use
the Ginzburg-Landau parameter
$\kappa=\lambda_{ab}/\xi_{ab}=$29$\pm$7 of LiFeAs from a SANS
experiment,~\cite{Inosov} the upper critical field
$\mu_0H_{c2}^{\parallel c}(=\sqrt{2}\kappa B_c(0))$ is
9.4$\pm2.3$~T. For comparison, $\mu_0H_{c2}^{\parallel c}$ from
penetration measurements is 17~T.~\cite{Cho}

\begin{figure}[tbp]
\includegraphics[width=7.5cm]{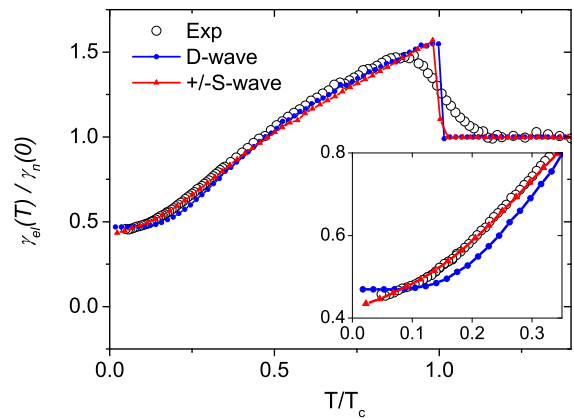}
\caption{(color online) Normalized electronic specific heat
coefficient (black open circles) and the best fits of the two SC gap models: d-wave (blue circles) and $\pm$s-wave (red
triangles) models. For the d-wave model, $2 \Delta_{max} /T_c =
3.5$ and $\Gamma= n_{imp}/ \pi N_{normal} =0.13$ are obtained for the best results. Here, $\Delta_{max}$ is the maximum gap of the d-wave gap $\Delta
(\theta)$. For the $\pm$s-wave model, $|\Delta|_{small} /
|\Delta|_{large}=0.46$, $2 \Delta_{large} /T_c =3.8$, and $\Gamma=
n_{imp}/ \pi N_{total} =0.17$ are obtained from the least squares fit. We note that the inversely related ratios between $|\Delta|_{small} / |\Delta|_{large}$ and
$N_{small}/N_{large}$ are an intrinsic property of the $\pm$s-wave
pairing model with a dominant interband paring interaction.~\cite{bang09} Inset: Low- temperature specific heat in the main panel is magnified to show a notable difference between the two gap models.}
\label{fig2}
\end{figure}
Figure 2 shows the electronic specific heat coefficient of LiFeAs
at zero magnetic field after subtracting the phononic
contribution: $\gamma_{el}(T)=C_{el}/T$=$(C-C_{ph})/T$. The
specific heat jump $\Delta C$ at $T_c$ is
7.65~mJ/mol$\cdot$K$^2$ at zero field. The jump relative to the
normal state specific heat $C_n$ is $~0.5$ (=$\Delta C/C_n$),
which is much smaller than the BCS value 1.43, indicating that the
quasiparticles participating in the SC condensation experience
strong elastic scattering because inelastic scattering usually
enhances the jump ratio.~\cite{bang04} Also the substantial value
of the Sommerfeld coefficient $\gamma_0$ in the SC state, which
accounts for about 45~\% of the normal state value $\gamma_n(0)$,
also supports the presence of strong (unitary) scatterers because
weak (Born limit) scatterers are not capable of inducing states inside the SC gap of unconventional superconductors.~\cite{bang09} These results suggest that any
model to explain the nature of the SC gap symmetry of LiFeAs
should take into account the effects of impurity scattering.

To extract information on the gap symmetry from the specific
heat of LiFeAs, we consider two typical unconventional SC
gap models that represent two extreme limits, i.e., a d-wave SC gap
with nodes and a $\pm$s-wave gap without nodes, and fit the
experimental data in Fig.2. Here we used the d-wave model to represent the generic behavior of a nodal SC gap state. The impurity effect is calculated
using the self-consistent $\mathcal{T}$-matrix approximation
(SCTA).~\cite{bang09} In each gap model, we determine the fitting
parameters such as $2 \Delta/ T_c$, the impurity concentration
parameter $\Gamma=n_{imp}/\pi N(0)$ by finding the best
overall fit to the data, where $N(0)$ is the zero-energy density of states. The results are shown in Fig.~2 overlayed with the experimental data. Both models provided reasonably good fittings for the overall shape, the specific heat jump $\Delta C$,
and the $\gamma_{el}(T=0~K)$ value. However, we find that the two models show
qualitative differences for $\gamma_{el}(T)$ at low temperatures and
this low temperature behavior of $\gamma_{el}(T)$ is what reflects the intrinsic properties of the gap symmetry, with which we can unambiguously identify the most compatible gap symmetry.

A pure d-wave superconductor is expected to produce a $T$-linear specific heat
coefficient such as  $\gamma(T) = \alpha T$ because of the
V-shaped density of state (DOS) $N(\omega) \sim \alpha^{'}
\omega$. However, the large value of $\gamma(T=0)$ ($\sim
0.45 \gamma_{n}$) indicates a substantial amount of unitary
impurities. When the unitary impurities create a zero energy impurity band inside the d-wave gap, the DOS around zero energy becomes a constant, $N(\omega)
\sim N_0$, which in turn produces a constant, temperature
independent, specific heat coefficient $\gamma(T)$ for a
finite range of low temperatures, as can be seen in the
inset of Fig.2 (blue circles). Our experimental
data, however, show $\gamma(T)\sim \gamma_{0} + \alpha T$, implying that the low energy DOS should have the form $N(\omega) \sim N_0 +
\alpha^{'} \omega$. This type of constant $+$ V-shape
DOS was indeed predicted for a $\pm$s-wave gap with unitary
impurities.~\cite{bang09} As can be seen in Fig.2 and its inset,
the $\pm$s-wave gap model with unitary impurities provide an
excellent fit at low temperatures as well as for the overall shape
of $\gamma_{el}(T)$, but the d-wave model fails to explain the low temperature behavior.

\begin{figure}[tbp]
\includegraphics[width=7.5cm]{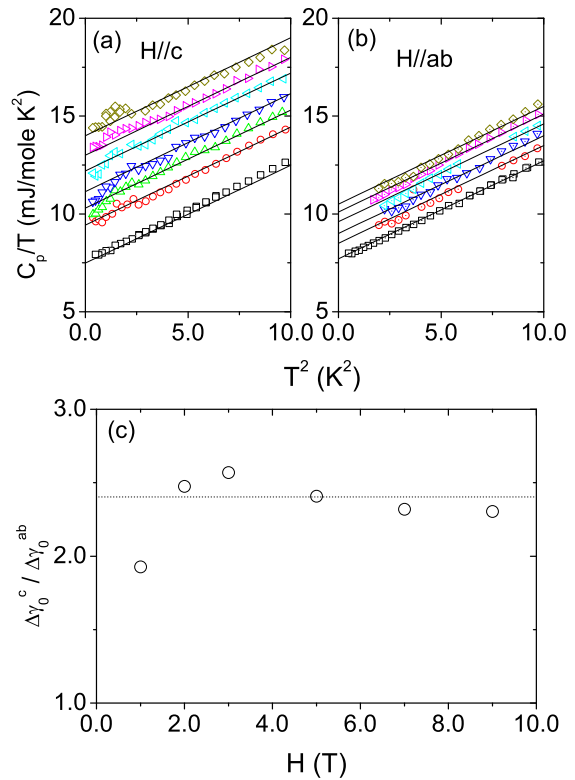}
\caption{(color online) Low temperature specific heat
capacity for magnetic fields applied parallel to the crystalline $c$-axis and
$ab$-plane in panels (a) and (b), respectively. In Fig.~3(a) the magnetic field intensity is 0, 1, 2, 3, 5, 7, and 9~T from bottom to top curves, and it is 0, 1, 3, 5, 7, and 9~T in Fig.~3(b). Straight lines are the results from the least-squares fit of $C/T=\gamma(H)+AT^2$ and used to obtain the zero temperature value of $C$/$T$.
(c) Anisotropy of the Sommerfeld coefficient which is obtained from the extrapolation of the data in Fig.~3(a) and (b) for the two field orientations. Here $\Delta \gamma_0^{c}=\gamma_0^c(H)-\gamma_0^c(0 T)$ for magnetic field along the c-axis.}
\label{fig3}
\end{figure}
The dependence on magetic field of the low-temperature specific heat
of LiFeAs is shown in Fig.~3(a) and~3(b) for different field
orientations of $H$$\parallel$$c$ and $H$$\perp$$c$, respectively.
The lack of a Schottky-like anomaly up to the highest applied magnetic
field enables us to unambiguously determine the Sommerfeld
coefficient $\gamma(H)$ from the least-squares fits of
$C(H)/T=\gamma(H)+AT^2$ to the low$-T$ specific heat. As shown in Fig.~3(c), the anisotropy in the quasiparticle density of states $(\gamma_0^{c}(H)-\gamma_0)/(\gamma_0^{ab}(H)-\gamma_0)$ is almost
constant ($\approx$2.4) over the experimental field range and an anisotropy
ratio of the upper critical fields between the two field
directions is estimated to be 1.7$\pm0.2$ through the
relationship:~\cite{Jang,Ichioka}
$\delta\gamma_0^c/\delta\gamma_0^{ab}$=$(0.5/0.3)(H_{c2}^{ab}/H_{c2}^c)^{0.7}$.
The $H_{c2}$ anisotropy from this thermodynamic measurement is in
good agreement with that obtained from transport properties where
$H_{c2}$ was directly measured under high magnetic
field.~\cite{Cho, khim11}

A sub-linear magnetic field dependence of $\gamma_0 (H)$ is
displayed in Fig.~4, which is often taken as evidence for a
nodal gap structure such as a d-wave state that is known to have a
generic $\sqrt{H}$ dependence due to Doppler effects of the nodal
quasiparticles by the supercurrent circulating around the
vortices. Recent theoretical work, however, showed that the
$\pm$s-wave state with different gap sizes ($\Delta_{small}\neq \Delta_{large}$) can also show a strong field dependence of $\gamma(H) \propto \sqrt{H}-H$.~\cite{bang10} In
Fig.~4, we show theoretical calculations for both $\pm$s-wave
(dotted line) and d-wave (dashed line) models to fit the $\gamma_0
(H)$ data with $H$$\parallel$$c$, where we use the same fitting
parameters obtained from the analysis of the specific heat as a
function of temperature (see Fig.~2). It is clear that the $\pm$s-wave model equally well reproduces the field dependence of $\gamma (H)$ which has been taken as a signature for a nodal gap structure. Thus, in combination with our analysis of the zero field temperature dependence, our data favors the sign changing multiband superconductivity of
$\pm$s-wave state. This multiple SC gap scenario is consistent with recent ARPES,~\cite{stockert11} penetration depth,~\cite{kim11} and heat transport measurements,~\cite{tanatar11} where two distinct full SC gaps were reported. 
\begin{figure}[tbp]
\includegraphics[width=8cm]{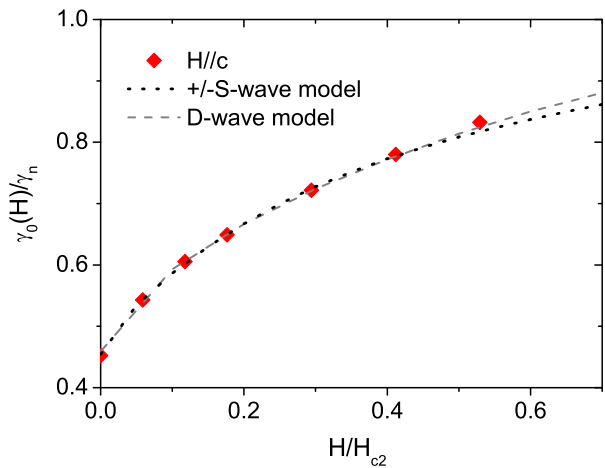}
\caption{(color online) Field dependent Sommerfeld coefficient
extrapolated from Fig.~3a for field applied along the crystalline c-axis. The black dotted line is the theoretical fit with the $\pm$s-wave model and grey dashed line is the fit with the d-wave model, respectively. Here we used the upper critical field $H_{c2}$ of 17~Tesla from Ref.~[29].}
\label{fig4}
\end{figure}

To summarize, we have reported the specific heat capacity of a LiFeAs
single crystal under magnetic field. The small specific heat jump
ratio at $T_c$ and substantial residual density of states at
$T=0$~K in the SC state indicate that impurity effects due to
strong unitary scatterers should be included to explain the
specific heat results. The zero-field specific heat data at high
temperatures above 2~K and sub-linear magnetic field dependence of
the Sommerfeld coefficient are equally well described by nodal and
nodeless multiband SC gaps with impurity effects. However, the
low-temperature electronic specific heat below 2~K excludes the
nodal d-wave state, but is consistent with the sign changing
multiband superconductivity of $\pm$s-wave state.

This work was supported by the NRF grant funded by Korea government (MEST) (No.~2010-0029136 \& 2011-0021645). Work at Los Alamos was performed under the auspices of the U. S. Department of Energy/Office of Science and supported in part by
the Los Alamos LDRD program. YB was supported by the grant
NRF-2010-0009523 funded by the National Research Foundation of
Korea. DJ acknowledges support from the Postdoctoral Research Program of Sungkyunkwan University (2011).

\end{document}